\documentclass[preprint]{aastex}
\usepackage{epsfig}
\usepackage{wasysym}
\usepackage{color} 
\def\ascnode {\mbox{\wasyfamily\char19}} 
 
\def\saturnring1  {\mbox{\wasyfamily\char21}}  
\def\vernal {\mbox{\wasyfamily\char23}}

\def\saturnring2 {\mbox{\wasyfamily\char31}}

\def\earth {\mbox{\wasyfamily\char38}}

\shorttitle{{\it{HST/COS}} Spectroscopy of BB Dor} 
\shortauthors{Godon et al.}
\begin{document}
\bibliographystyle{apj}

\title{Hubble Space Telescope Cosmic Origins Spectrograph Spectroscopy
of the Southern Nova-like BB Doradus in an Intermediate State       
}

\author{Patrick Godon\altaffilmark{1}, Edward M. Sion} 
\affil{Department of Astrophysics \& Planetary Science, 
Villanova University, Villanova, PA 19085}
\email{patrick.godon@villanova.edu ; edward.sion@villanova.edu}

\author{Boris T. G\"ansicke} 
\affil{University of Warwick, Department of Physics, 
West Midlands, CV4 7AL, UK}  
\email{boris.gaensicke@warwick.ac.uk}

\author{Ivan Hubeny}
\affil{Steward Observatory, 
The University of Arizona, Tucson, AZ 85721, USA} 
\email{hubeny@as.arizona.edu}

\author{Domitilla de Martino}
\affil{INAF aLS Osservatorio Astronomico di Capodimonte, 
Napoli, I-80131, Italy} 
\email{demartino@na.astro.it}

\author{Anna F. Pala}
\affil{Department of Physics, 
University of Warwick, Conventry, CV4 7AL, UK} 
\email{A.F.Pala@warwick.ac.uk}

\author{Pablo Rodr\'iguez-Gil}
\affil{Instituto de Astrof\'isica de Canarias, V\'ia L\'actea, 
s/n, La Laguna, E-38205 Santa Cruz de Tenerife, Spain  \\ 
\&  \\  
Departmento de Astrof\'isica, Universidad de La Laguna, 
La Laguna, E-38206 Tenerife, Spain}  
\email{prguez@iac.es}

\author{Paula Szkody}
\affil{Department of Astronomy, University of Washington,
Seattle, WA 98195}
\email{szkody@astro.washington.edu}

\author{Odette Toloza}
\affil{Department of Physics, 
University of Warwick, Conventry, CV4 7AL, UK} 
\email{O.F.C.Toloza@warwick.ac.uk}

\altaffiltext{1}
{Visiting at the Henry A. Rowland Department of Physics and 
Astronomy, The Johns Hopkins University, Baltimore, MD 21218}

\begin{abstract}

We present a spectral analysis of the  
Hubble Space Telescope Cosmic Origins Spectrograph 
spectrum of the southern VY Scl nova-like variable BB Doradus,
obtained as part of a Cycle 20 {\it HST/COS} 
survey of accreting white dwarfs in cataclysmic variables. 

BB Dor was observed with {\it COS} during 
an intermediate state with a low mass accretion rate,
thereby allowing an estimate of the white dwarf temperature. 
The results of our spectral analysis show that the white dwarf
is a significant far ultraviolet component with  a temperature 
of $\sim$35,000-$\sim$50,000$~$K, 
assuming a $0.80~M_{\odot}$ WD mass ($\log(g)=8.4$).  
The disk with a mass accretion rate of 
$\approx 10^{-10}~M_{\odot}~$yr$^{-1}$ contributes about 1/5 to 1/2  
of the far ultraviolet flux.

\end{abstract}


\section{Introduction}

\subsection{Accretion Rates and WD Temperatures in CVs}

Cataclysmic variables (CVs) are close binaries in which 
a white dwarf (WD) accretes matter from 
a low-mass companion donor 
star by  means of either an accretion disk around
the WD, or an accretion column - when the WD has a strong magnetic
field \citep{war95}. 
When enough material has been accreted
onto the surface of the WD, every $\sim 10^3-10^5$ yr or so, the hydrogen-rich
envelop undergoes a thermonuclear runaway - the classical nova explosion
\citep{sta71}.

In the standard model \citep{rap82,rap83},  
CVs are believed to evolve from long to short binary orbital period ($P$)  
driven by angular momentum losses (AMLs).        
Magnetic stellar wind braking is the dominant AML mechanism
at long periods, and it shuts down  around $P \sim 3$ hr, 
as the secondary becomes fully convective. Since the mass-loss rate
is suddenly reduced, the secondary contracts and the system becomes 
a detached binary, continuing to evolve to shorter periods ($P<3$ hr)      
driven now by gravitational wave emission as the dominant AML mechanism.
Around $P \sim 2$ hr, the mass transfer restarts as the Roche lobe 
remakes contact with the donor. The system then re-appears as 
active and continues to evolve reducing its period. 
The majority of CVs in our galaxy should have already evolved  
to a period minimun near 80 min and now have degenerate brown dwarf-like
secondaries \citep{how01,nel16}. However, some discrepancies between the
observations and the standard model have arisen, prompting the suggestions
of additional (or alternative) theoretical scenarios 
(see e.g. \citet{kni11} and references therein for a review of the problem).  

In order to differentiate between the different possible theoretical paths, 
one needs to know
the distribution of the time-averaged mass accretion rate ($< \dot{M}>$) of CVs
as a function of the binary orbital period $P$ \citep{tow05}. 
Alternatively, the distribution
of the WD effective surface temperature $T_{\rm wd}$  versus $P$
can also be used \citep{tow09}.  
Therefore, assessing the mass accretion and/or WD effective temperature
of CV systems has become a primary goal.   

The mass accretion rate can be obtained by fitting theoretical 
accretion disk spectra to far ultraviolet (FUV) spectra of 
CV systems accreting a high rate, 
and the WD temperature is obtained by fitting synthetic 
photospheric spectra to FUV spectra of CV systems in which the accretion
is greatly reduced.  

The nova-like variables (NLs) form a class of non- or weakly-magnetic CVs characterized by their 
high mass accretion rate in which the accretion disk is the dominant    
source of ultraviolet and optical light \citep{lad90,lad91}. 
These systems are often observed for the
study of the accretion disk and to determine the mass accretion rate 
$\dot{M}$, since the WD 
cannot be detected, as it is outshone by the brightness of the accretion
structures.  This  
prevents determination of the WD parameters.  The VY Scl 
sub-class of NLs consists of systems that occasionally drop into a low 
brightness state \citep{hac93}, 
thereby allowing the study of the WD and its companion. 
Because of that, the VY Scl sub-class of 
NLs are of special interest, as they provide
the opportunity to assess the effective surface temperature of 
their WDs. So far, only a handful of NL systems have been studied
in the low state, revealing hot WDs with a temperature
$\sim$30,000 K to $\sim$50,000 K  \citep{tow09,rod15}.  
BB Doradus is a member of this sub-class of NL systems and is, therefore,
of special interest.

\subsection{The VY Scl Nova-Like Cataclysmic Variable BB Doradus} 

BB Doradus, also known as EC 05287-5847, was spectroscopically identified as 
a CV from the Edingburgh-Cape Survey \citep{che01}. 
It  was not brighter than $m_v \sim 16.5$
until November 1992, after which it went into a   
bright state reaching $M_{\rm v}\sim 14.6-13.6$. 
Since then, it has been observed to
exhibit occasional low states \citep{rod12,sch12}.  
A 45 d photometric coverage by  \citet{pat02} revealed 
super-humps and a photometric modulation of 0.14923$~$d (or 3.58$~$h).
More recently however, a radial velocity study,  
using for the first time the H$\alpha$ radial velocities during 
the low state  \citep{rod12}, 
put the orbital period at $0.154095 \pm 0.000003$ d ($3.69828 \pm 0.00007$ h), 
as expected for VY Scl stars which usually have orbital periods between
3-4\,h. 
\citet{rod12} suggest that the modulation of 0.14923 d derived by \citet{pat02}
may be identified as a negative superhump.
BB Dor also exhibits quasi-periodic oscillations (QPOs) 
with timescales between 878-1154 s \citep{che01}.  
Due to its very narrow
Balmer lines (with wings extending only $\sim 650~$km$~$s$^{-1}$ from
the line center), BB Dor is suspected to be a low-inclination system
\citep{che01}. 
The Galactic reddening value in the direction of BB Dor, as retrieved from  
the online GALEX MAST archive, is very small 
E(B-V)=0.035, therefore, we assume that the reddening of BB Dor 
is negligible E(B-V)$ \approx 0$. 

We observed BB Dor on 2007 July 09     
with the Far Ultraviolet Spectroscopic Explorer ({\it FUSE}), 
just days before the fatal failure of the reaction wheel of the telescope
\citep{god08},  
and found it in a high state with a visual magnitude $\sim 13.6-14.6$,   
revealing a mass accretion rate of $\sim 10^{-9}~M_{\odot}~$yr$^{-1}$ (for  
a standard  $\log(g)=8.4$ WD), and a distance estimate of 
$\sim$665 pc \citep{god08}. 
Taking into account the contribution of a hot 
($T_{\rm eff}=32,000~$K) WD in addition
to the disk, pushed the distance to $\sim 700~$pc, 
but overall, the inclusion of the WD did not significantly
improve the fit as the hot WD contributed only 10\% of the flux.

More recently, \citet{pal16} observed BB Dor with the 
Hubble Space Telescope Cosmic Origins Spectrograph ({\it HST/COS}),  
as part of a 122 orbit {\it HST/COS} program 
to measure the WD effective temperature in CVs in the low state 
or quiescence.  
In that analysis, \citet{pal16} modeled the {\it COS} spectra of CVs 
using synthetic spectra of the WD plus a simple power law or 
black body second component 
but did not include the contribution from an accretion disk. 
All the systems were observed
when the disk contribution to the FUV flux was expected to be negligible. 
BB Dor, however, was observed in an intermediate state
(see below), and  
the broad hydrogen Ly$\alpha$ absorption profile of the WD  
was not detected in its {\it COS} spectrum,  
and for that reason it was not modeled    
by \citet{pal16}. 

In the present work, we extend the  spectral analysis of the
{\it COS} spectrum of BB Dor to also include the disk component; 
namely, our spectral modeling includes synthetic spectra of 
a WD with an accretion disk. 
We also take into account the results of the {\it FUSE} spectroscopic 
analysis \citep{god08}.  
By doing so, we aim to obtain constraints on both the WD effective 
surface temperature and the mass accretion rate of BB Dor during its 
intermediate state. 

\section{The {\it COS} Spectrum of BB Dor} 

Data from the AAVSO showed that BB Dor dropped from its usual high 
state ($m_{\rm v} \approx 15$)  into a low state 
in mid-January 2013 reaching a magnitude of $\sim 19 - 19.5$,
thereby triggering the HST/{\it COS} observations of BB Dor. 
However, the system luminosity started to increase  on Jan 19,  
and as a consequence, at the time of the {\it COS} observations 
(2013 February 13, $\sim$25 days after its deep minimum),   
BB Dor was found in an intermediate state, with a magnitude of $17.5-18.0$.
It took another $\sim$25 days to return to its high state. 

The {\it COS} instrument was set with the detector in the 
FUV channel configuration,  
with the G140L grating centered at a wavelength of (CENWAVE=) 1105 \AA\ ,
covering the wavelength range 1110 \AA - 2150 \AA ,  
with a resolution of $R \approx 3000$.  
The photons were collected through the Primary Science Aperture (PSA) with  
a diameter of 2.5 arcsec. 
The {\it COS} data consist 
of a total of 7272$~$s of good exposure in TIME-TAG mode.  
We used the 4 FP-POS positions to improve S/N and minimize
the effects of flat-field artifacts. 
The data (LC1VVD010) consisted of 5 sub-exposures, 
as one of the FP-POS positions spread over two consecutive 
spacecraft orbits. 
A final spectrum was then  
constructed by co-adding the 5 X1D files generated by 
CALCOS (v3.1; \citet{hod11,hod07}).  

The longer wavelengths beyond 2000 \AA\ 
were found to be very noisy and that portion of the spectrum had to  
be discarded. 
Consequently, the final spectrum starts at 1117 \AA\ 
and ends at 2000 \AA .  
The spectrum is presented in Fig.1.

Strong emission lines from low and high ionization species 
have been identified and are annotated in Fig.1. The 1180 \AA\ - 1250 \AA\ 
wavelength region does not show any discernible hydrogen 
Ly$\alpha$ profile/wings 
(from the WD and/or accretion disk) due to the many 
emission features in that region. 
In fact, the spectrum is contaminated 
with strong and broad geocoronal emission of 
Ly$\alpha$ (1216 \AA\ ) and O\,{\sc i} (1302 \AA\ ). 
In the longer wavelengths, 
$\sim 1850-1900$ \AA\  ,  
we identify emission lines from Al\,{\sc iii} (1857 \& 1862 \AA\ ), 
and there seems  to be additional emission features beyond 1900$~$\AA\ 
that remained unidentified.  Because of this,  
we ignore the portion of the spectrum in the longer wavelengths
($ \lambda > 1850$ \AA ) in our modeling (our spectral modeling
does not include emission lines).

Of all the emission lines from the source, the C\,{\sc iv} line 
is the strongest and best resolved. We, therefore, use it to 
assess the inclination of the system as this line is believed to 
be forming in the hotest region in the disk, 
namely, the very inner disk $r \sim 1-2R_{\rm wd}$. 
The C\,{\sc iv} (1u) line is made of a doublet with rest wavelengths at  
1548.20 \AA\ and 1550.78 \AA . The modeling of the doublet with a double 
Gaussian gives a FWHM of 6.85 \AA\ (for each Gaussian),  and twice as much 
for the maximum broadening at their base. This translates 
into a velocity of $\sim 650~$km$~$s$^{-1}$
for the FWHM, and $\sim$1300$~$km$~$s$^{-1}$ for the maximum broadening. 
Assuming that the maximum
broadening is due to the largest Keplerian speed in the disk, 
obtained at $r\sim R_{\rm wd}$, one obtains
an inclination of $\sim$20$^{\circ}$ for the disk/binary for an average 
$0.8~M_{\odot}$ CV WD mass with a maximum Keplerian speed of 
$\sim$3900$~$km$~$s$^{-1}$.    
Assuming a larger radius for the region where the C\,{\sc iv} line forms,
$r \approx 2 R_{\rm wd}$, gives an inclination of $\sim$30$^{\circ}$.

There are also some absorption  lines, mostly at short wavelengths. 
All the absorption lines are much narrower than the emission lines. 
We present the lines in Fig.2. 
We tentatively identify the following lines:   
Si\,{\sc iv}  (1122.5 \& 1128.3),  
Si\,{\sc iii} (1140.6, $\sim$1142, \& $\sim$1145),  
C\,{\sc iii} ($\sim$1175),   
S\,{\sc iii} (1190.2, 1194.2, \& 1201),   
Si\,{\sc iii} (1206.5),  
Si\,{\sc ii} (1260),  
C\,{\sc ii} (1334.5 \& 1335.7),  
Si\,{\sc iii} ($\sim 1500$),  
and Si\,{\sc ii} (1526). 
The nitrogen line N\,{\sc i} (1134) is most probably terrestrial 
in origin and has often been observed in {\it FUSE} spectra of CVs. 
Since the absorption lines might potentially
originate in the WD photosphere or the accretion disk (when seen 
at low inclination), we  further discuss them in sections 4 \& 5.

A comparison of the continuum flux 
level of the {\it COS} spectrum to that of the archival {\it FUSE} 
spectrum (taken on 2007 July 9) shows that the {\it COS} continuum 
flux level is lower than the 
{\it FUSE} continuum flux level by a factor of 20. 
Namely, in the overlap region $\sim 1110-1185$ \AA\ 
the {\it FUSE} continuum reaches $\sim 2 \times 10^{-13}$ 
 erg$~$s$^{-1}$cm$^{-2}$ \AA$^{-1}$, 
while the {\it COS} cotinuum reaches $\sim 1 \times 10^{-14}$ 
 erg$~$s$^{-1}$cm$^{-2}$ \AA$^{-1}$. 
Though the system was not caught in a deep low state,  
this fortunate occurrence gives us the unique opportunity 
to study possible photospheric emission from the  
WD, as the mass accretion rate must have dropped substantially,   
thereby possibly exposing the WD. 

In order to assess the contribution of the WD to the total 
flux, we use the TIME-TAG data to generate 3 light curves. 
A first light curve is generated by integrating all the data over 
the entire wavelength coverage except for the known airglow 
emission lines of Ly$\alpha$, O\,{\sc i}, and N\,{\sc i}; 
a second light curve is made with the C\,{\sc iv} line
(1535-1564\AA );
and a third light curve is made using the continuum in the 
range 1424-1523\AA .  
In Fig.3 we show the 3 light curves, with each normalized to one.  
The continuum varies in the same way as the C\,{\sc iv} emission, which 
we know does not come from the white dwarf. Therefore, if the continuum is
the sum of the disk and white dwarf, the fact that the variability 
($\sim 20$\% of the total flux) is produced by the disk, 
indicates that the white dwarf cannot contribute more than 
$\sim 80$\% of the flux. 

In our analysis we model the COS spectrum 
of BB Dor with WD models, disk models and 
composite models including the contribution from both  
the WD photosphere and an accretion disk. 
Our main goal is to derive the WD effective surface temperature
and the mass accretion rate of the system, to add one data 
point to the handfull of NL systems with known $T_{\rm eff}$.  
Our spectral analysis technique is described in the next section. 

\section{Synthetic Spectral Modeling \& Analysis} 

In order to derive  
the mass accretion rate $\dot{M}$ and the WD temperature $T_{\rm eff}$,  
we carry out a quantitative comparison (a ``fit'') of the observed spectrum 
to theoretical disk and WD spectra for a wide range of parameters.  

The synthetic WD stellar atmosphere model spectra are generated  
using the FORTRAN suite of codes TLUSTY, SYNSPEC and ROTIN \citep{hub88,hub95}.
The basic input parameters are the stellar effective surface temperature, 
surface gravity and chemical composition.  
In a first step, a converging stellar atmospheric structure is computed 
by successive runs of TLUSTY. 
The TLUSTY model atmospheres have hydrogen and helium taken explicitly,
i.e. their selected levels
are treated in NLTE, and the bound-bound and bound-free transitions
between these levels determine the total opacity and emissivity.
Carbon, nitrogen, and oxygen are treated implicitly, i.e. they only
contribute to the particle and charge conservation, assuming LTE.
Next, a run of the spectral synthesis code
SYNSPEC, using the output model atmosphere structure from TLUSTY as input
is done. 
In this step, all opacity sources, i.e., the lines of all other elements
are considered (unless it is specifically required not to).
The short code, ROTIN, is then used to account for the projected stellar 
rotational velocity (and/or instrumental) broadening, including the 
effect of limb darkening.   

In this manner, we generate WD stellar photospheric models with effective
temperatures ranging from $\sim$15,000 K to $\sim$60,000 K in increments of 
500 K to 1,000 K. 
The NLTE treatment is turned on for the hot WD models
($T_{\rm eff}>35,000$ K), using the 
approximate NLTE treatment of lines option in SYNSPEC \citep{hub95}.  
We chose here a value of $\log(g)=8.4$ which matches 
a WD mass of $0.80~M_{\odot}$, a rounded value of the average WD mass in CVs 
of $0.83 \pm 0.23 ~M_{\odot}$ \citep{zor11}.  
Since the spectrum has emission lines and thus a disk component,
we use only solar composition (set up in SYNSPEC) as it would be impossible to derive 
chemical abundances for such a spectrum. 
We vary the
stellar rotational velocity $V_{\rm rot} \sin(i)$ from $100~$km$~$s$^{-1}$
to $500~$km$~$s$^{-1}$ in steps of $50~$km$~$s$^{-1}$.  
The WD rotation ($V_{\rm rot} \sin(i)$) rate is determined by fitting the
WD model to the spectrum while paying careful attention to the 
line profiles.  
The mass radius relation from \citet{woo90},  
for carbon-oxygen and non-zero temperature WDs, is 
used to obtain the radius of the WD. 

The accretion disk spectra were generated   
assuming that the disk is made of a collection
of annuli, where each annulus has a temperature $T(r)$ and
gravity $\log(g(r))$ given by the standard disk model \citep{sha73,pri81},  
for a given central mass $M_{\rm wd}$ and accretion rate $\dot{M}$. 
For each annulus, an atmosphere model was generated using TLUSTY, 
followed by a run of SYNSPEC to obtain an annulus spectrum. 
The contribution of all the annuli were 
combined together using the code DISKSYN, and a final disk spectrum was obtained
for any given inclination angle (we assume the disk to lay in the 
orbital plane of the binary). 
In the present work we use the public grid of disk models 
from \citet{wad98}, in which a detailed explanation of the entire numerical
procedure is given. 

In our analysis, we use a grid of models of synthetic spectra of WDs and accretion disks
covering a wide range of values of the WD temperature $T_{\rm eff}$, gravity
$\log(g)$, projected rotational velocity $V_{\rm rot} \sin(i)$, inclination
$i$, mass accretion rate $\dot{M}$ and abundances. 
The grid of disk models has the following inclination angles: 
$i= 18^{\circ}$, $41^{\circ}$, $60^{\circ}$, $75^{\circ}$, and $81^{\circ}$. 
Since the inclination of the system is low \citep{che01,sch12}, 
and based on our assessment in the previous section, we use $i=18^{\circ}$, 
the lowest value in the grid of models. 
Furthermore, as the flux level in the {\it COS} spectrum is 20 times 
lower than in the {\it FUSE} spectrum, 
we expect our {\it COS} spectral analysis to reveal  
a mass accretion rate lower than that derived 
from the {\it FUSE} spectrum, namely  
$\dot{M}_{COS} < 10^{-9}~M_{\odot}~$yr$^{-1}$. 

The distance to the system is not known accurately, but the results 
of the {\it FUSE} analysis implied a distance d$< 1$ kpc \citep{god08}.  
\citet{rod12} estimated a distance of d $\sim 1-2$ kpc based on 
a WD temperature of $30,000$ K and an M3-M4 secondary spectral type.   
The spectrum exhibits few absorption lines, and, as stated earlier, we decided 
to use solar abundances in our modeling. With our assumption of a
standard CV WD mass of $0.8~M_{\odot}$, we keep $\log(g)$ constant at $8.4$.  
Therefore, the parameters that we vary in our model fits are 
the temperature of the WD, its rotation rate, and the mass accretion rate
of the disk. 

Before carrying out a synthetic spectral fit of the spectra,
we mask portions of the spectrum with emission lines, since our synthetic 
spectral code does not generate emission lines.

After having generated grids of models for the {\it COS}
spectrum of BB Dor,  
we use FIT \citep{numrec}, a $\chi^2$ minimization routine,
to compute the reduced $\chi^{2}_{\nu}$ 
($\chi^2$ per number of degrees of freedom $\nu$) 
and scale factor (which gives the distance) for each model fit.  
While we use a $\chi^2$ minimization technique, we do not 
blindly select the least $\chi^2$ models, but we examine the models 
that best fit some of the spectral features and that are most consistent with the values
of the system parameters.

\section{Results}

\subsection{Single WD modeling} 

Despite the facts that the hydrogen Ly$\alpha$ absorption feature 
is not detected in the spectrum 
\citep{pal16}, and that the WD cannot contribute more than $\sim$80\% of 
the flux, we decide to first model the spectrum with a single 
WD alone. Such a modeling will confirm that a WD alone is not enough
to account for the observed flux and will help us identify the lines
forming in the WD photosphere (if any).  

We first try to fit the slope of the continuum, and we find that it 
agrees well with a 23,000 K WD
(our model \#1, see Table 1),
but the Ly$\alpha$ region of the model forms a large broad absorption 
feature not seen in the observed spectrum. 
The distance obtained for this model 
is also far too short, only 291 pc, and the $\chi^2_{\nu}=3.07$.  

As we increase the temperature, we find a better agreement in the Ly$\alpha$
region, but the model becomes too blue as its 
slope becomes steeper. The distance also increases,  
and it reaches $\sim$1 kpc for a 50,000 K WD (model \#6), 
with $\chi^2_{\nu}=1.94$. 

The least $\chi^2$ model is obtained for 
a 30,000 K WD (model \#3), with a distance of 515 pc. 
This model is shown in Fig.4. 
Even for this model the  Ly$\alpha$
region is also too broad and too deep and the continuum is too steep.  
It is likely that the Ly$\alpha$ region (1170-1250\AA ) is entirely
dominated by broad emission lines.  
We note that setting the stellar rotational velocity 
to 150$~$km$~$s$^{-1}$ in this model gives a reasonable fit to the Si\,{\sc ii} 1260 line. 
The short wavelength region is shown in detail with
line identification in Fig.5. 
However, the Si\,{\sc ii} 1260 line is the only line
that can be matched with this WD model. The expected 
Si\,{\sc ii} 1251 \& 1265 lines, as seen in the theoretical spectrum,
are not observed. 
The C\,{\sc iii} (1175) line is contaminated
with broad emission, but its absorption feature on top of the emission
is too shallow compared to the theoretical spectrum.  

We find that we can fit some set of lines by changing the
WD temperature, but each temperature provides a fit to 
only a few lines. 
For example, in Fig.6, we set the WD temperature to $T=35,000$ K, 
with a projected rotational velocity of 300$~$km$~$s$^{-1}$,
and obtain a relatively good fit to the 
Si\,{\sc iv} (1122.5 \& 1128.3) and Si\,{\sc iii} (1140.6, 
1141.6, 1142.3, \& $\sim$1145) lines.
However, the Si\,{\sc ii} (1260) line is not reproduced in this model. 

Similarly, a WD temperature of $T=40,000$ K, 
with a projected rotational velocity of 150$~$km$~$s$^{-1}$ is needed to fit 
the C\,{\sc ii} 1324 \& $\sim$1335 absorption lines, 
however, none of the other lines were fitted. 

These are clear indications that the observed lines do not form 
in the stellar WD photosphere. 
Overall, the single WD model does not provide a satisfactory fit.
The results for the single WD model fits 
are recapitulated in the first part of Table 1 (models \#1-6).

\subsection{Single Disk Modeling}  

As the absorption lines cannot satisfactorily be fitted 
with single WD models, we first check whether the disk alone can
provide such a fit. We find that,
{\it in order to fit the narrow absorption lines in the {\it COS} spectrum,}
the inclination of the system  
would have to be extremely small ($\sim 3^{\circ}$). Namely, 
the disk would have to be almost face-on to minimize the
broadening of the lines from the Keplerian velocity in the disk.  
Since there is no
indication that this is the case, we deduce that the lines do not
form in the disk either. 
The width of the C\,{\sc iv} emission line 
implies an inclination of $\sim 20^{\circ}$,   
we therefore set the  inclination to $i=18^{\circ}$ (the smallest value in our 
grid of disk models). We vary the mass accretion rate
from $\dot{M}=10^{-8.0}~M_{\odot}$yr$^{-1}$ 
down to $\dot{M}=10^{-10.5}~M_{\odot}$yr$^{-1}$ 
in logarithmic steps of 0.5 (models \#7-12 in Table 1, second part). 
Here, we do not make any more attempt to fit the absorption lines,
but rather we concentrate on the continuum.  

The best fit is for $\dot{M}=10^{-8.5}~M_{\odot}$yr$^{-1}$ 
(model \#8), with $\chi^2_{\nu}=1.32$,
but it gives an unacceptably large distance (more than 4 kpc) 
and the spectral features 
in the shorter wavelengths are not reproduced by the disk model. 
This model is presented in Fig.7. 

The slope of the observed spectrum agrees better with 
a $\dot{M}=10^{-9.5}~M_{\odot}$yr$^{-1}$ 
disk model (\#10, Fig.8), which has a distance 
(1356 pc) in better agreement with the $\approx 1-2$ kpc. However, the 
$\chi^2_{\nu}$ value (1.87) of this model 
is significantly larger than for the   $\dot{M}=10^{-8.5}~M_{\odot}$yr$^{-1}$ 
model. In the vicinity of Ly$\alpha$ region ($\sim 1170-\sim 1250$) 
the fit is rather poor.  

Overall, the single disk model provides only a marginal improvement
over the single WD model. However, due to the limits on the distance to the
system, the mass accretion rate {\it in the single disk model}  
is constrained to be in the range 
$10^{-10.5} M_{\odot}$/yr $< \dot{M} < 10^{-9}M_{\odot}$/yr.  

\subsection{Combined WD + Disk Modeling}  

Since neither the single white dwarf nor the single disk models fit
the spectrum adequately, we next attempt models with both components. 
We run each single disk model in combination with a WD
with a temperature varying from 20,000 K to 60,000 K, in steps of 1000 K, 
a total of almost 200 
model fits. We present here only the best fit models. 

Since the disk models with a large mass accretion rate gave 
an unacceptably large distance, we present here only the lower mass
accretion rate disk + WD models. For completeness we list some
large mass accretion rate disk + WD models in table 1    
(models \#13-18). 

As the mass accretion rate is set to $\dot{M}=10^{-9.5}~M_{\odot}$/yr
the disk model is improved with the addition of a hot WD.
In Fig.9 we present such a model with $T_{\rm wd}=35,000$ K, giving a distance
of 1.5 kpc, and where the WD contributes $\approx$1/5 of the FUV flux
and the disk contributes the remaining 4/5. 
The projected rotational velocity of the WD is  
$V_{\rm rot} \sin{i} =200~$km$~$s$^{-1}$. 
This model (\# 20 in Table 1) has a $\chi^2=1.49$.  
As the WD temperature is further increased,
to $T_{\rm wd}=48,000$ K (model \#21) and larger  (60,000 K,  model \#22),  
the model provides a small improvement with $\chi^2_{\nu} \approx 1.3$. 

The fit can be further improved to $\chi^2_{\nu}\approx 1.2$ by decreasing
further the mass accretion rate: $\dot{M}=10^{-10}~M_{\odot}$yr$^{-1}$  
with $T_{\rm wd}=40-50,000$ K. Models \#24-26 
are all very similar and we present model \#24 in Fig.10. These models give 
a distance of the order of 1-1.2 kpc.  

As we further decrease $\dot{M}$ to $\dot{M}=10^{-10.5}~M_{\odot}$yr$^{-1}$  
($\approx 3 \times 10^{-11}~M_{\odot}$yr$^{-1}$ ), 
the lowest $\chi^2_{\nu}$ (1.23) for that mass accretion rate is obtained
with $T_{\rm wd}=35,000$ K - model \#29.
This model is presented in Fig.11, it gives a distance of 761 pc. 
However, this model has a WD contributing more than 80\% of the 
{\it COS} flux and, therefore, it has to be rejected. 
Other models with  $\dot{M} = 10^{-10.5}~M_{\odot}$yr$^{-1}$ 
have either a distance that is too short (when $T_{\rm wd}$ is decreased) 
or have a WD contributing more than 80\% of the flux 
(when $T_{\rm wd}$ is increased).    

The best fits obtained for the combined WD + disk components have a 
mass accretion rate  
$ \dot{M} = 10^{-9.5} -10^{-10}~M_{\odot}$yr$^{-1}$ 
and WD temperature $T_{\rm wd} \sim 35,000$ K and larger.

\section{Discussion and Conclusion} 

We found that the absorption lines in the COS spectrum of BB Dor 
do not match the WD photosphere lines, and cannot originate in the disk. 
Absorption lines were also observed in the FUSE spectrum of BB Dor
taken during its high state, when the disk dominated the FUV. 
Because of that the FUSE absorption lines could not originate 
in the WD photosphere, and, because of their width, 
they could only originate in the disk 
if $i < 10^{\circ}$ \citep{god08}. Since the inclination is probably 
larger, i.e. $\sim 20^{\circ}$, we conclude that the absorption lines
observed in the FUV spectra of BB Dor 
{\it during high and intermediate states}  
must come from material either above the disk and WD,
or from circumbinary material, or from the interstellar medium (ISM).

The ISM lines in the FUSE spectra of CVs are usually very narrow
with a broadening $<< 1$ \AA ,
consisting of low ionization species such as 
Fe\,{\sc ii}, S\,{\sc i}, C\,{\sc ii}, Si\,{\sc ii}, and Ar\,{\sc i} 
\citep{god06,god12}, while the absorption features  
in the {\it COS} spectrum of BB Dor have a larger velocity broadening  
$> 1$ \AA . Nonetheless, broader ISM lines have been observed in the
FUV spectra of some CVs \citep{mau88}, which enable us to tentatively 
identify the following lines as being interstellar in origin: 
Si\,{\sc ii} (1190.42, 1193.29), N\,{\sc i} (1199.55, 1200.22, 1200.71), 
Si\,{\sc iii} (1206.51), S\,{\sc ii} (1250.59, 1253.81, 1259.52), 
Si\,{\sc ii} (1260.42), \& C\,{\sc ii} (1334.53, 1335.70).
 
No hydrogen Ly$\alpha$ absorption profile was discernible due to the 
broad emission in the spectral region 1170-1250\AA . 
Other regions of the spectrum 
also exhibit some broad and strong emission lines.  
As a consequence, and also due to the unknown mass and undertain
distance estimate, 
the modeling of the COS spectrum of BB Dor proved to be challenging.  

In our modeling we assumed an inclination $i=18^{\circ}$, a standard 
CV WD mass of $0.8~M_{\odot}$, solar abundances and an upper limit
of 2 kpc for the distance \citep{rod12}. 
The combined WD plus accretion disk models with a moderately low 
mass accretion rate disk and a hot WD 
appear to be the best fits as far as the lowest $\chi^2_{\nu}$ value is concerned.   

These results have to be further considered while taking into account 
that the system was in an intermediate state with a flux 20 
times lower than during the {\it FUSE} observation when the system was in its  
high state \citep{god08}. 
The {\it FUSE} spectrum of BB Dor had a 
continuum flux level decreasing below 1000 \AA\  
implying that the mass accretion rate 
{\it had to be} $\approx 10^{-9}~M_{\odot}$yr$^{-1}$ 
at the time of the {\it FUSE} observation, 
as models  
with larger mass accretion rate ($\dot{M} > 10^{-9}M_{\odot}~$yr$^{-1}$)
were hotter and contributed more flux in the shorter wavelengths
($<1000$ \AA ) of {\it FUSE}. 
The {\it FUSE} spectral fit also gave a distance 
estimate of $\approx 700$ pc.  
At the time of the COS observation, BB Dor must have had  a lower mass 
accretion rate, namely   $\dot{M} \le 10^{-9.5}~M_{\odot}$yr$^{-1}$
(i.e. 3 times lower than during the {\it FUSE} observations), 
and more likely
$\dot{M} \approx 10^{-10}~M_{\odot}$yr$^{-1}$ 
(i.e. 10 times lower than during the {\it FUSE} observations). 

As the WD cannot contribute more than a maximum of 
$\sim$80\% of the continuum, which would be achieved in 
the extreme case where the disk
luminosity itself varies by 100\%, 
we reject models \#29 \& \#30, and due
to the distance constraint models \#27 \& \#28 have also to be rejected.  
In other words, during the COS observations, the mass accretion rate   
must have been of the order of $\dot{M}=10^{-10}~M_{\odot}$yr$^{-1}$ or larger.  

We therefore conclude that BB Dor, during its COS observation, 
most probably had 
a mass accretion rate $\dot{M} \sim 10^{-10}~M_{\odot}$yr$^{-1}$,  
and a WD temperature of $\sim$35,000 to $\sim$50,000 K,
as presented in Fig.10. 
This is typical of NL VY Scl systems, and is 
higher than expected for CVs just above the period gap.  
We note, however, that the distance obtained from the fit changes dramatically
in the solutions, while its estimate is poorly constrained
(from $\sim 700$ pc to about 2 kpc). Anticipated Gaia data will soon 
provide distances to many CVs, and should help resolve the best
solution for BB Dor and many other CVs.  
A short distance ($< 1$ kpc) would imply a low mass accretion rate 
($ < 10^{-10}M_{\odot}~$yr$^{-1}$) with a WD temperature 
of $\sim 35,000$ K,
while a large distance ($\sim 2$ kpc) would imply larger mass 
accretion rate $( > 10^{-10}M_{\odot}~$yr$^{-1} )$ 
and WD temperature  ($T \sim 50,000$ K). 

Only a few VY Scl NL systems have been observed in an intermediate state
and the modeling of their FUV spectra has shown mixed results. 
AC Cnc seems to reveal a disk and WD, though the mass accretion rate
and WD temperature depend strongly on the unknown distance \citep{bis12}. 
BZ Cam has a FUV spectrum consistent with a 12,500 K black body 
\citep{pri00}, but it is a unique peculiar object exhibiting an 
emission bow-shock nebula \citep{ell84}. MV Lyr was one of the few VY Scl 
systems
that were modeled in a low state \citep{hoa04,god12},
but its intermediate state
spectrum could not be modeled with a standard disk model and part of
the disk had to be set to a low isothermal temperature \citep{lin05}.  
However, these mixed results are characteristic of NL systems in high state
as well. While some NLs have been successfully modeled with standard
disk spectra (e.g. 
IX Vel, V3885 Sgr, V794 Aql, RZ Gru; \citet{lin07,lin09,god07,bis12}) 
others appear impossible to model with standard disk spectra 
(e.g. 
RW Sex, V751 Cyg, V380 Oph, UX UMa, QU Car; \citet{lin10,zel09,lin08b,lin08a}). 
The results of the 
statistical study of \citet{pue07} suggest that the temperature profile
of the standard disk model be revised at least 
in the innermost parts of the
disk in order to properly model some of the disk-dominated CV systems.  
With, so far, no existing alternative disk model than the standard 
disk model, the current results for BB Dor are at least consistent with
typical WD temperatures in VY Scl NLs, and improved disk models, as 
well as accurate distances obtained by Gaia in the near future, will 
provide better constraints on the physical parameters of this system.  
 
\clearpage 

\acknowledgments

We would like to thank Knox Long and Christian Knigge for reading
and commenting on an early version of this manuscript.  
P.G. wishes to thank Bill Blair for his kind hospitality 
at the Rowland Department of Physics 
and Astronomy in the Johns Hopkins University, Baltimore, Maryland. 
Special thanks to the Austral Variable Star Observer Network (AVSON),
and the American Association of Variable Star Observers (AAVSO) 
for their monitoring of BB Dor.
This research is part of the GO program 12870 
based on observations made with the NASA/ESA 
Hubble Space Telescope, obtained at and funded by the Space Telescope Science
Institute, which is operated by the Association of Universities for 
Research in Astronomy Inc., under NASA contract NAS 5-26555. 
The research leading to these results has received funding from the
European Research Council under the European Union's Seventh 
Framework Programme (FP/2007-2013) / ERC Grant Agreement n.320964 (WDTracer).

\begin{table} 
\caption{Results of Synthetic Spectral Modeling} 
\begin{tabular}{ccrccrr}
\hline
model & $T_{wd}$ & $Log(\dot{M})$ & WD/disk &  d      &  $\chi^2_{\nu}$ & Fig. \\  
number & ($10^3$K) & $M_{\odot}$/yr &         & $<pc>$  &                 &      \\  
\hline
1 &   23.0    & ---             &  WD    &  291    &    3.07         &   --- \\ 
2 &   27.0    & ---             &  WD    &  412    &    1.98         &   --- \\ 
3 &   30.0    & ---             &  WD    &  515    &    1.92         &    4  \\ 
4 &   35.0    & ---             &  WD    &  692    &    1.94         &   5-6 \\ 
5 &   40.0    & ---             &  WD    &  864    &    1.98         &   --- \\ 
6 &   50.0    & ---             &  WD    & 1052    &    2.17         &   --- \\ 
\hline
7 &   ---     & -8.0            & disk   & 6808    &    1.36         &   ---  \\ 
8 &   ---     & -8.5            & disk   & 4288    &    1.32         &    7   \\ 
9 &   ---     & -9.0            & disk   & 2545    &    1.36         &   ---  \\ 
10 &  ---     & -9.5            & disk   & 1356    &    1.87         &    8   \\ 
11 &  ---     & -10.0           & disk   &  679    &    4.66         &   ---  \\ 
12 &  ---     & -10.5           & disk   &  322    &   12.14         &   ---  \\ 
\hline
13 &  30.0    & -8.0            & 0.6/99.4& 6828   &   1.36          &   ---  \\ 
14 &  50.0    & -8.0            & 2.4/97.6& 6890   &   1.37          &   ---  \\ 
15 &  30.0    & -8.5            & 1.5/98.5& 4319   &   1.32          &   ---  \\ 
16 &  50.0    & -8.5            & 6/94    & 4415   &   1.33          &   ---  \\ 
17 &  30.0    & -9.0            & 4/96    & 2598   &   1.35          &   ---  \\ 
18 &  50.0    & -9.0            & 15/85   & 2754   &   1.34          &   ---  \\ 
19 &  25.0    & -9.5            & 6/94    & 1400   &   1.85          &   ---  \\ 
20 &  35.0    & -9.5            & 21/79   & 1524   &   1.49          &    9   \\ 
21 &  48.0    & -9.5            & 37/63   & 1693   &   1.31          &   ---  \\ 
22 &  60.0    & -9.5            & 46/54   & 1817   &   1.31          &   ---  \\ 
23 &  35.0    & -10.0           & 53/47   &  967   &   1.47          &   ---  \\ 
24 &  40.0    & -10.0           & 65/35   & 1094   &   1.21          &    10  \\ 
25 &  45.0    & -10.0           & 70/30   & 1175   &   1.19          &   ---  \\ 
26 &  50.0    & -10.0           & 74/26   & 1244   &   1.23          &   ---  \\ 
27 &  27.0    & -10.5           & 66/34   &  516   &   2.79          &   ---  \\ 
28 &  30.0    & -10.5           & 76/24   &  602   &   1.68          &   ---  \\ 
29 &  35.0    & -10.5           & 86/14   &  761   &   1.23          &   11   \\ 
30 &  40.0    & -10.5           & 91/9    &  920   &   1.32          &   ---  \\ 
\hline
\end{tabular} 
\end{table}

\begin{figure}
\plotone{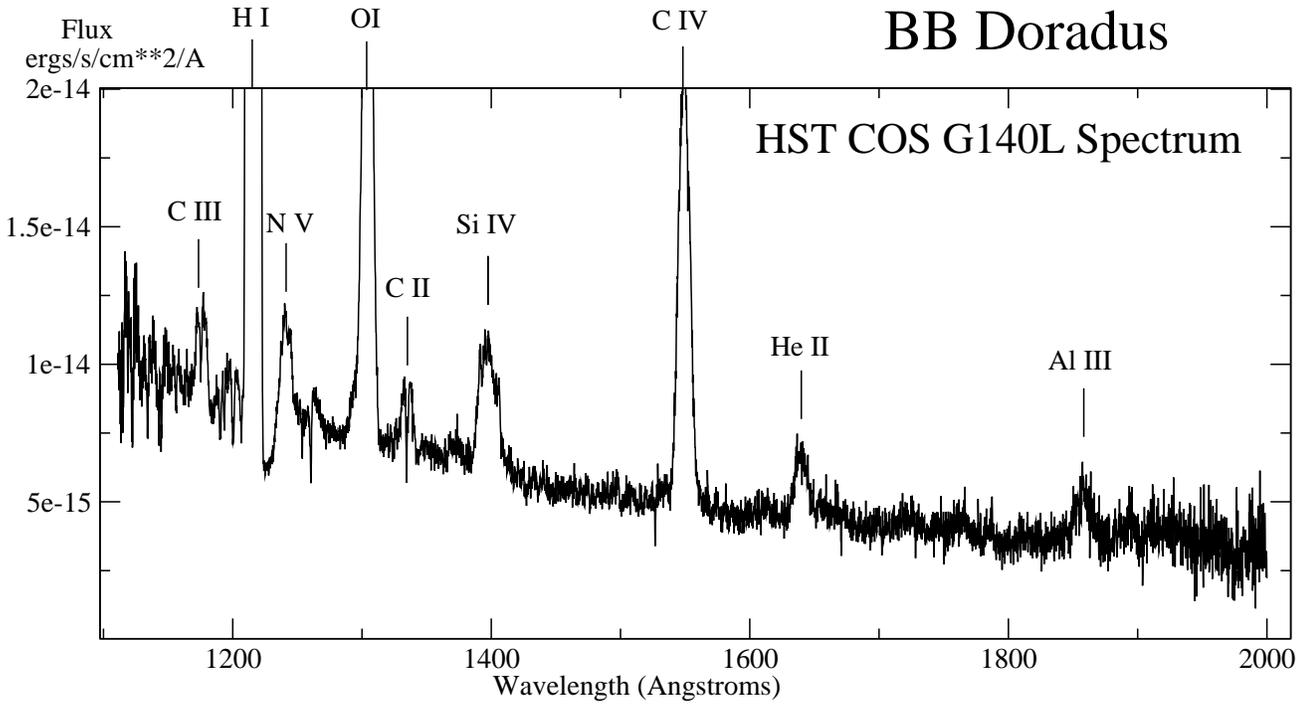}             
\vspace{-7.cm} 
\figcaption{The {\it HST/COS} Spectrum of BB Dor is shown with 
line identification. 
The spectrum presents some broad and 
strong emission lines of high ionization species
such as C\,{\sc iii} (1175), N\,{\sc v} (1240), 
Si\,{\sc iv} (1400), C\,{\sc iv} (1550), and also   
of lower ionization species such as 
C\,{\sc ii} (1335),   and He\,{\sc ii} (1640).  
The hydrogen Lyman$\alpha$ (1216) and O\,{\sc i} (1302) features 
are due to geocoronal contamination. 
Toward the longer wavelengths, after 
1700 \AA\ and especially after 1900 \AA\ ,  there are additional 
weak but broad emission features that remain unidentified.  
The region $ \lambda > 2000$ \AA\ is dominated by noise and has 
been discarded for clarity. 
For that reason in the modeling we limit the fitting to $ \lambda < 1850 $\AA\ . 
}
\end{figure}

\clearpage 

\begin{figure}
\vspace{-5.cm} 
\plotone{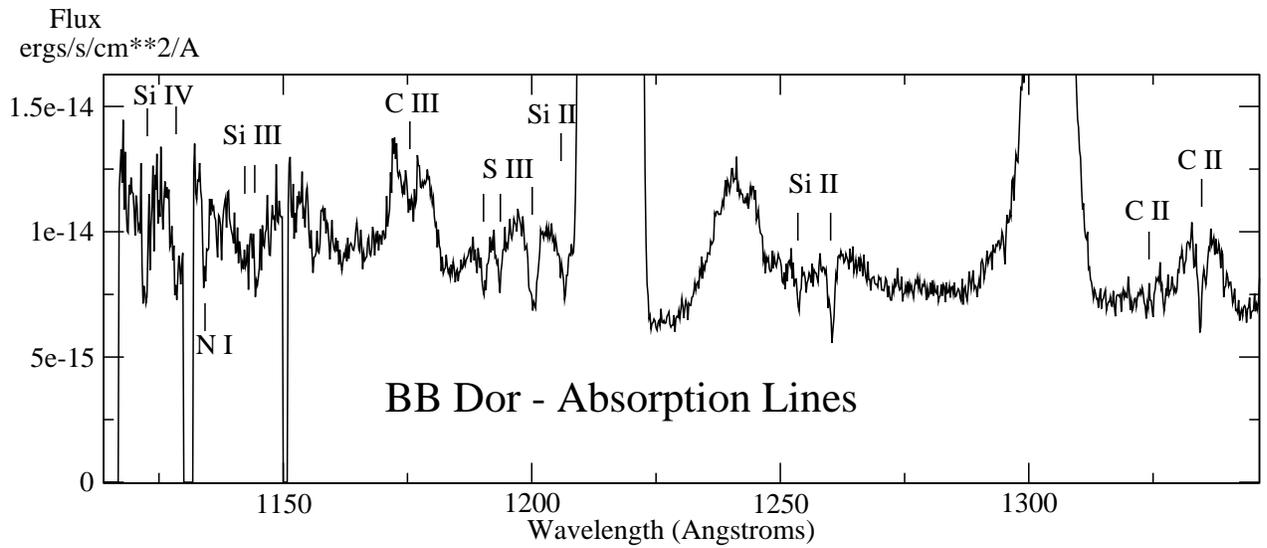}                 
\vspace{-5.cm} 
\caption{Absorption lines are tentatively identified and annotated
in the shorter wavelength region of the COS spectrum of BB Dor.
The nitrogen line N\,{\sc i} (1134) is possibly geocoronal.  
The two narrow gaps ($\sim 1130$ \& $\sim 1150$ \AA ) 
are due to further post-processing of the spectrum,
namely, cutting off unreliable data in the sub-exposures 
to remove small detector artifacts 
that could be misidentified as absorption lines. 
}
\end{figure}

\clearpage 

\begin{figure}
\hspace{-2.cm}               
\epsscale{1.2}  
\includegraphics[width=20cm,height=6cm,angle=00]{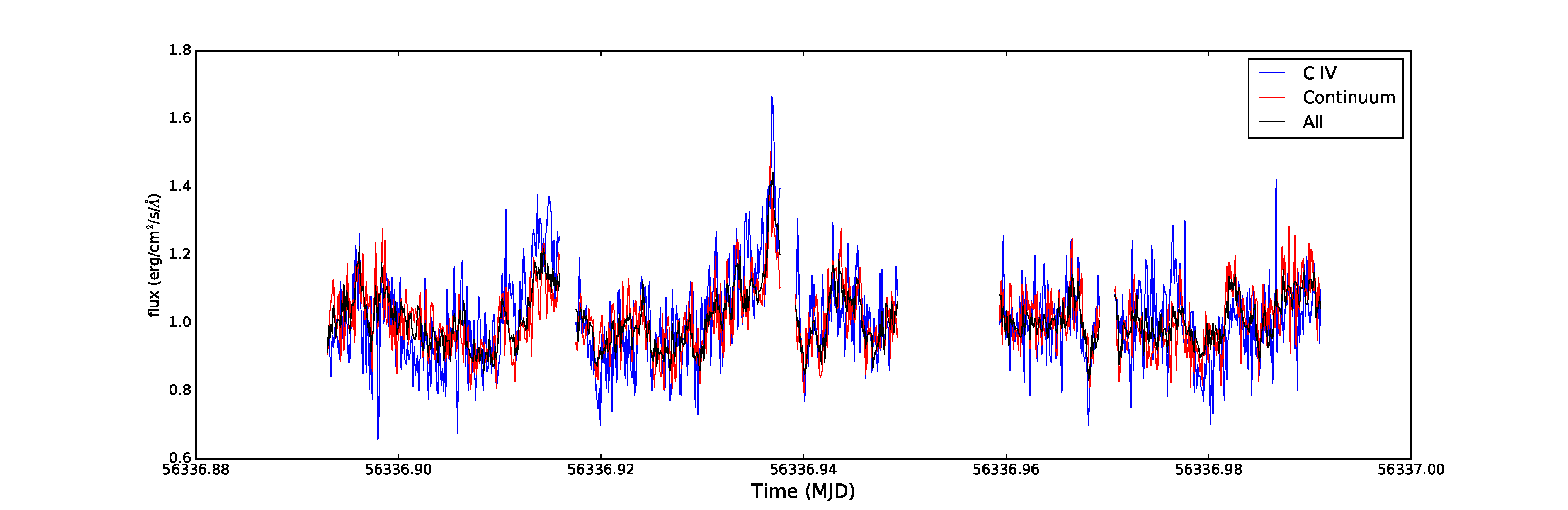} 
\caption{ 
Light curves are shown for (i) the entire spectral range of 
{\it COS} (in black), (ii) the carbon line only (1424-1523\AA ; in
blue), and (iii) for the continuum in the range 1535-1564\AA\ 
(in red). The 5 segments correspond to the 5 exposures. 
The flux has been normalized to one and the time is given using the  
modified Julian date.  
}
\end{figure}

\clearpage 

\begin{figure}
\epsscale{1.0}  
\plotone{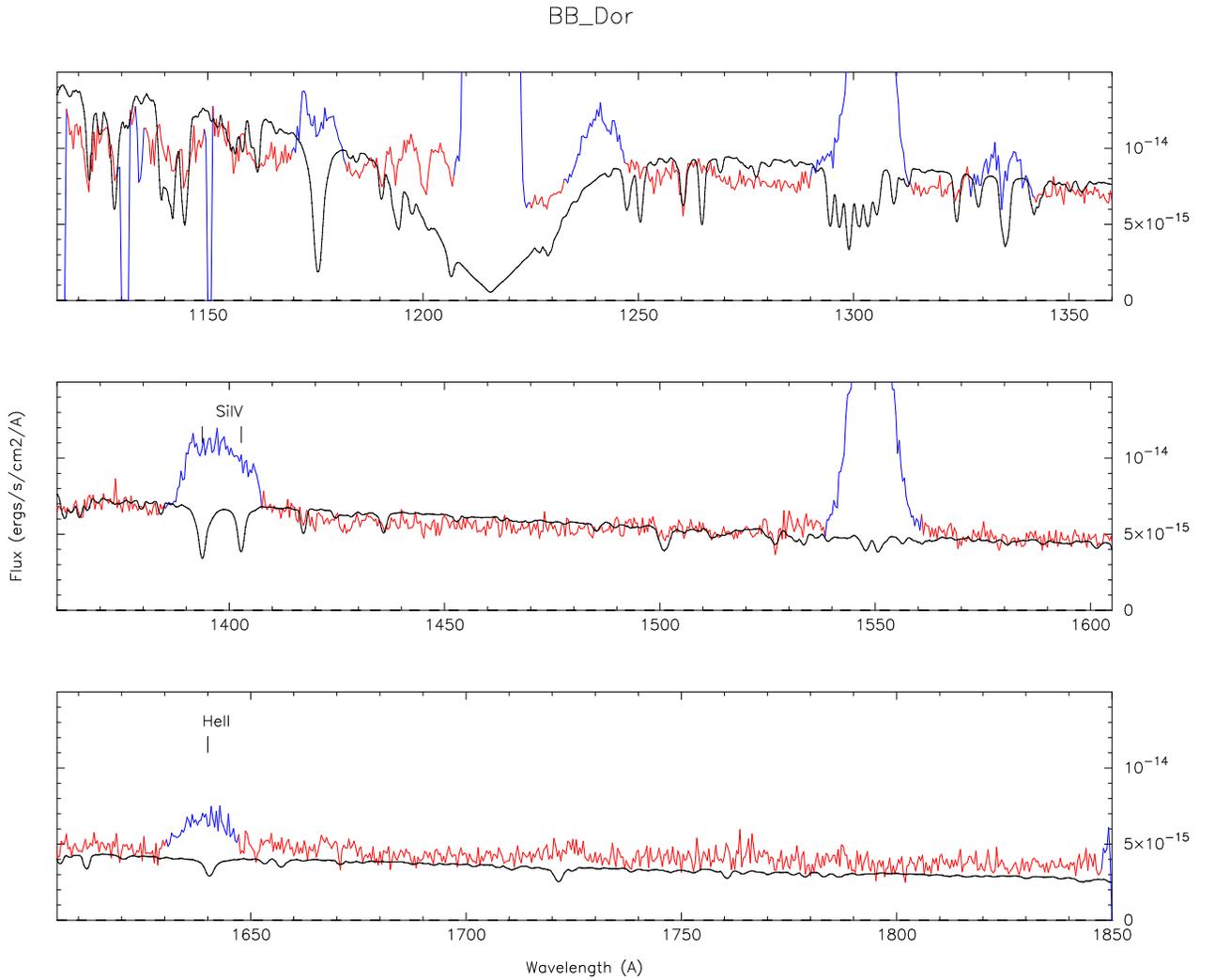} 
\caption{
A WD model fit (in black) to the COS spectrum of BB Dor (in red). 
The blue portions indicate the 
regions of the spectrum that have been masked due to the 
strong emission features. 
The WD has a temperature $T=30,000$ K, gravity $\log(g)=8.4$, 
solar composition and projected rotational velocity of 200$~$km$~$s$^{-1}$. 
This is the best-fit (least $\chi^2$) WD model. Its slope is slightly
too steep and its hydrogen 
Ly$\alpha$ profile does not fit the observed spectrum. 
It is likely that the Ly$\alpha$ region (1170-1250 \AA ) is entirely
dominated by broad emission lines.  
}
\end{figure}

\clearpage 

\begin{figure}
\plotone{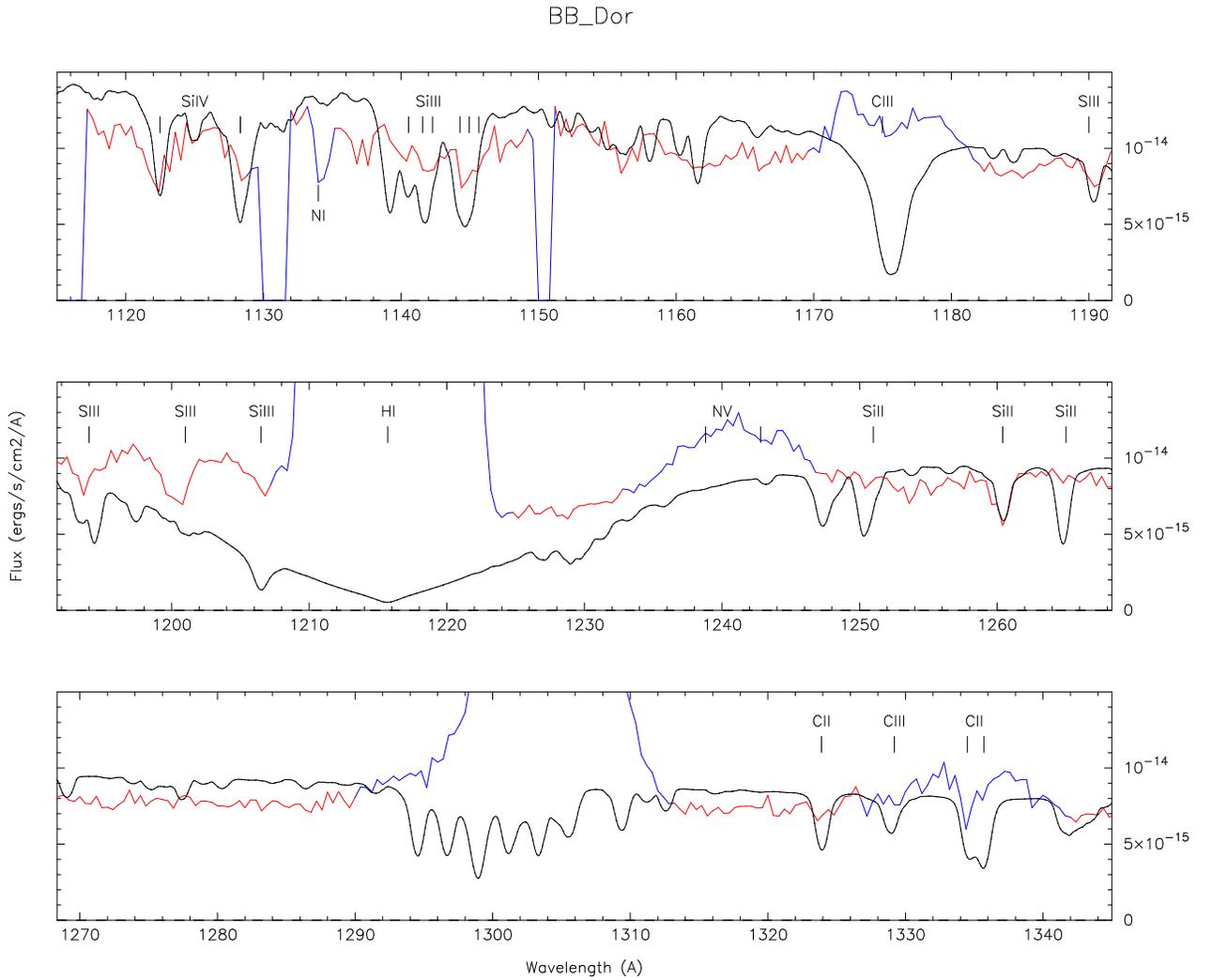}                  
\caption{
Same as in Fig.3, but with a stellar rotational velocity 
of 150$~$km$~$s$^{-1}$ to fit the Si\,{\sc ii} 1260 line. 
The short wavelength region is shown in details with
line identification. The Si\,{\sc ii} 1260 line is the only line
that can be matched with this WD model. The Si\,{\sc ii} 1251 \& 1265 
lines are not observed. 
The C\,{\sc iii} (1175) line is contaminated
with broad emission, but its absorption feature on top of the emission
is too shallow compared to the theoretical spectrum.  
These are clear indications that the observed lines do not form 
in the stellar WD photosphere. 
}
\end{figure} 

\clearpage 

\begin{figure}
\plotone{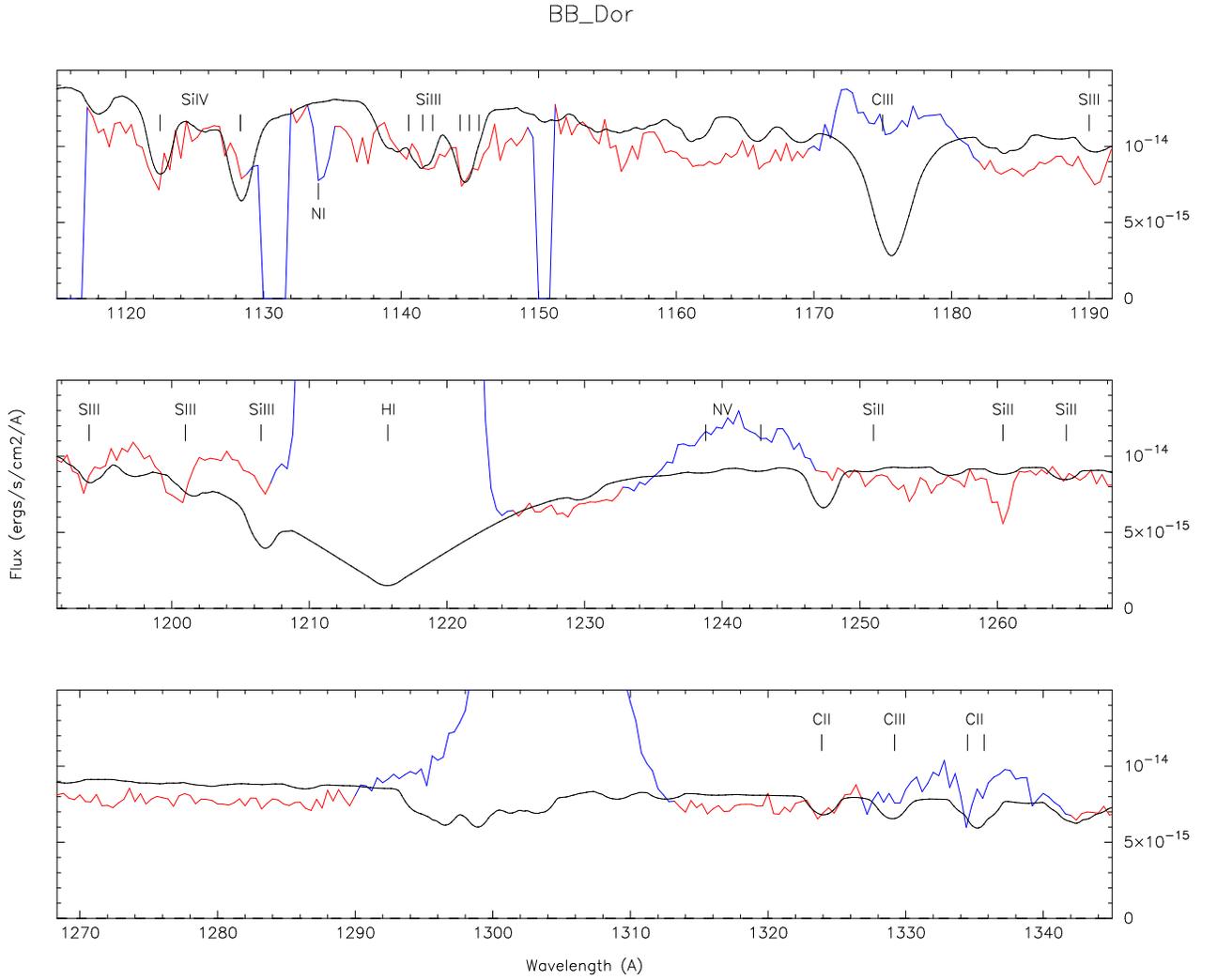}                  
\caption{
A WD model fit to the Si\,{\sc iv} \& Si\,{\sc iii} 
absorption lines in the very short wavelength region 
($\lambda < 1150$ \AA ) of the COS spectrum of BB Dor.  
The WD has a temperature $T=35,000$ K, gravity $\log(g)=8.4$, 
solar composition and projected rotational velocity of 300$~$km$~$s$^{-1}$.
The Si\,{\sc iv} (1122.5 \& 1128.3) and Si\,{\sc iii} (1140.6, 
1141.6, 1142.3, \& $\sim$1145) lines are relatively well fitted
but the Si\,{\sc ii} (1260) line is not reproduced.   
}
\end{figure}

\clearpage 

\begin{figure}
\plotone{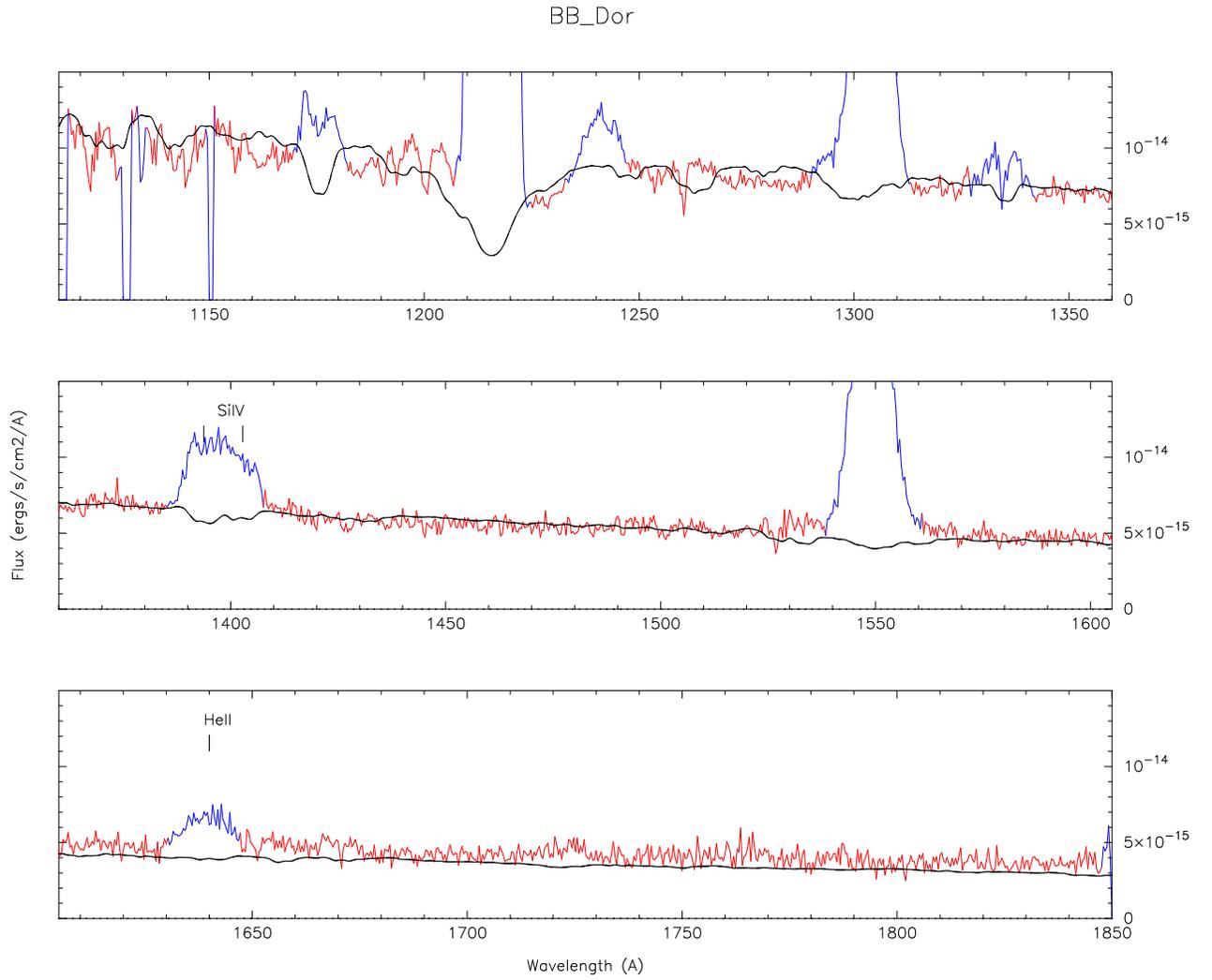}           
\caption{ 
A single disk model fit to the COS spectrum of BB Dor, with 
$M_{wd}=0.8~M_{\odot}$, $\dot{M}=10^{-8.5}~M_{\odot}$yr$^{-1}$, 
and $i=18^{\circ}$.
This is the best single disk model fit, but it gives a distance far too
large of 4288 pc.  
}
\end{figure}

\clearpage 

\begin{figure}
\plotone{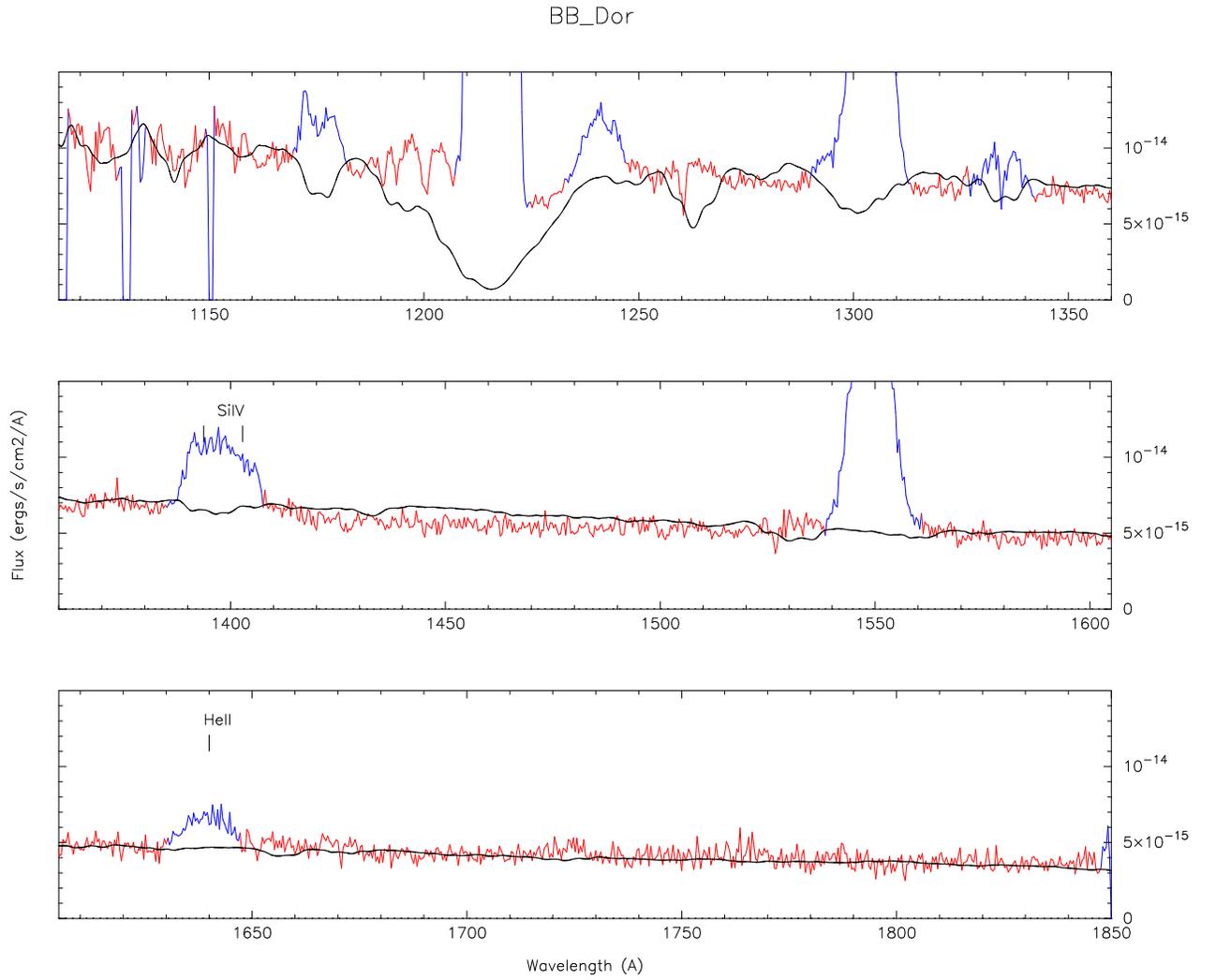}           
\caption{ 
A single disk model fit to the COS spectrum of BB Dor, with 
$M_{wd}=0.8~M_{\odot}$, $\dot{M}=10^{-9.5}~M_{\odot}$yr$^{-1}$, 
and $i=18^{\circ}$ 
giving a distance of 1356 pc.  
}
\end{figure}

\clearpage

\begin{figure}
\plotone{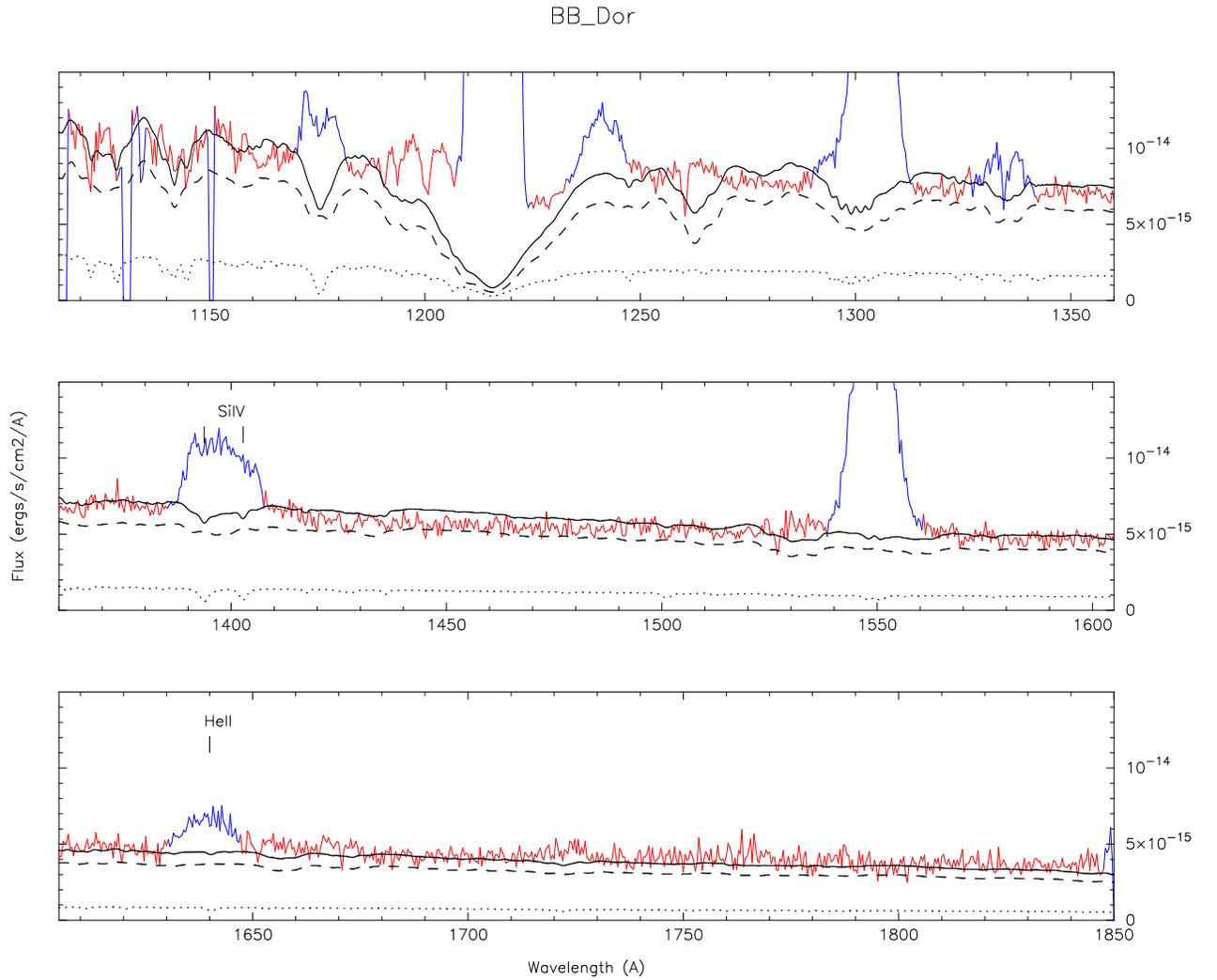}             
\figcaption{
The {\it COS} Spectrum of BB Dor 
has been modeled (solid black line) with a moderate mass accretion rate disk 
(dashed black line) plus a hot WD (dotted black line) to agree 
with the state in which it was observed. 
The WD contributes 21\% of the FUV flux, and the disk contributes
the remaining 79\%. 
Both the WD and the disk are assumed to have solar composition. 
The WD has a mass of $0.8~M_{\odot}$ and is accreting at a rate 
of $\dot{M}=10^{-9.5}~M_{\odot}$yr$^{-1}$  
$\approx 3 \times 10^{-10}~M_{\odot}$yr$^{-1}$,  
assuming an inclination of  $i=18^{\circ}$. 
For this moderate mass accretion rate, a very good fit is obtained 
as long as the WD has a temperature of $T=35,000$ K 
(shown here) or higher.  
The distance obtained from the fit is rather large $d=$1524 pc, 
and $\chi^2_{\nu}=1.49$. 
As the WD temperature is increased (up to $T \approx 60,000$ K), 
the fit is slightly  improved ($\chi^2_{\nu} \approx 1.3$),
but the distance approaches its 2 kpc upper limit.  
}
\end{figure}

\clearpage

\begin{figure}
\plotone{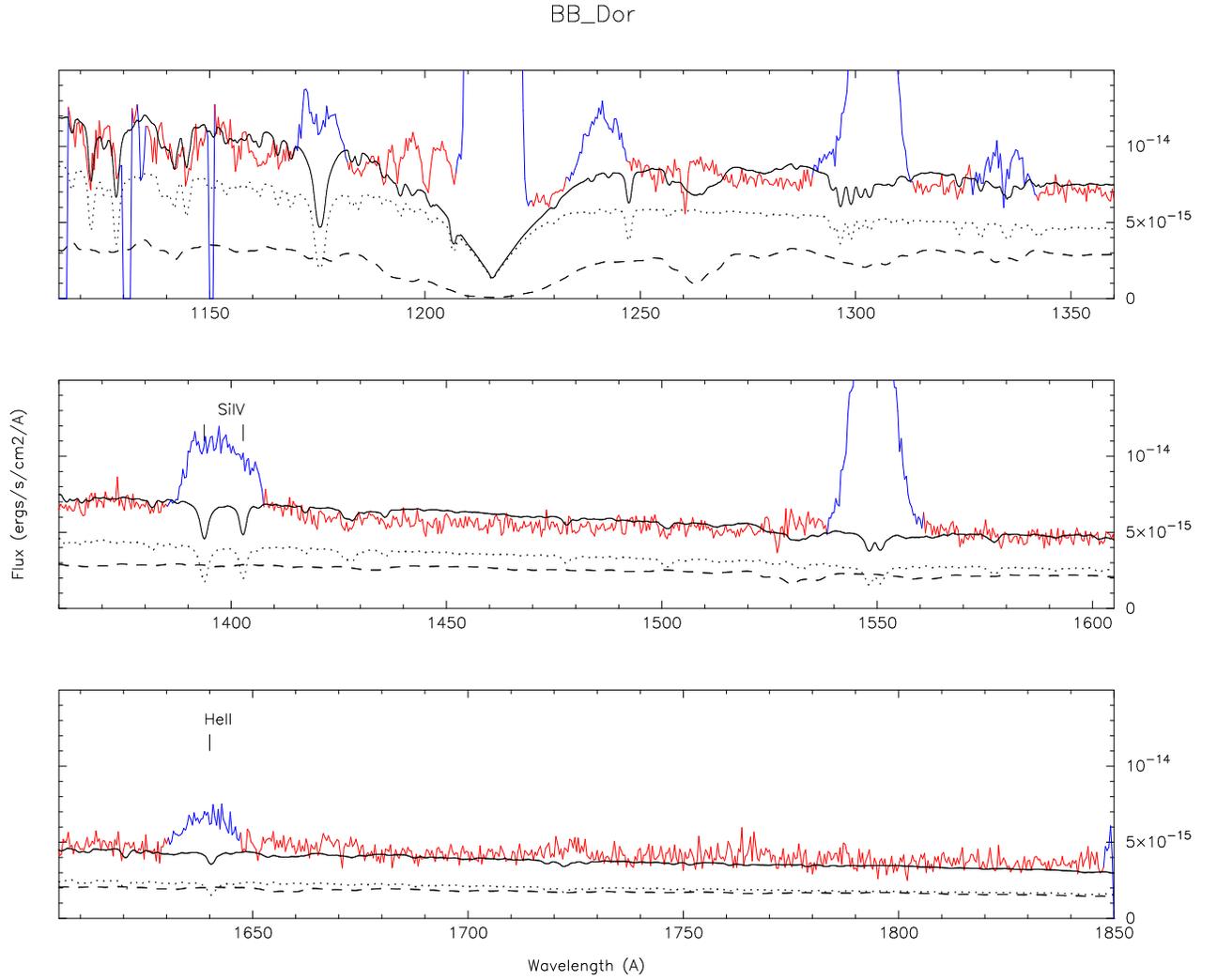}             
\figcaption{
Combined WD plus disk model to the COS spectrum of BB Dor. 
The WD has a temperature $T=40,000$ K, a projected
rotational velocity $V_{\rm rot} \sin{i} =200~$km$~$s$^{-1}$, 
and mass of $0.8~M_{\odot}$. 
The mass accretion rate is $\dot{M}=1 \times 10^{-10}~M_{\odot}$yr$^{-1}$  
with  $i=18^{\circ}$. 
The distance obtained is $d=$1094 pc, and the reduced $\chi^2_{\nu}=1.21$. 
The WD contributes 65\% of the FUV flux, and the disk contributes
the remaining 35\%. Here too, as long as the WD temperature is large
(up to $T\sim 50,000$ K), the fit remains very good 
($\chi^2_{\nu} = 1.2$), and the distance increases to 1.2 kpc. 
}
\end{figure}

\clearpage 

\begin{figure}
\plotone{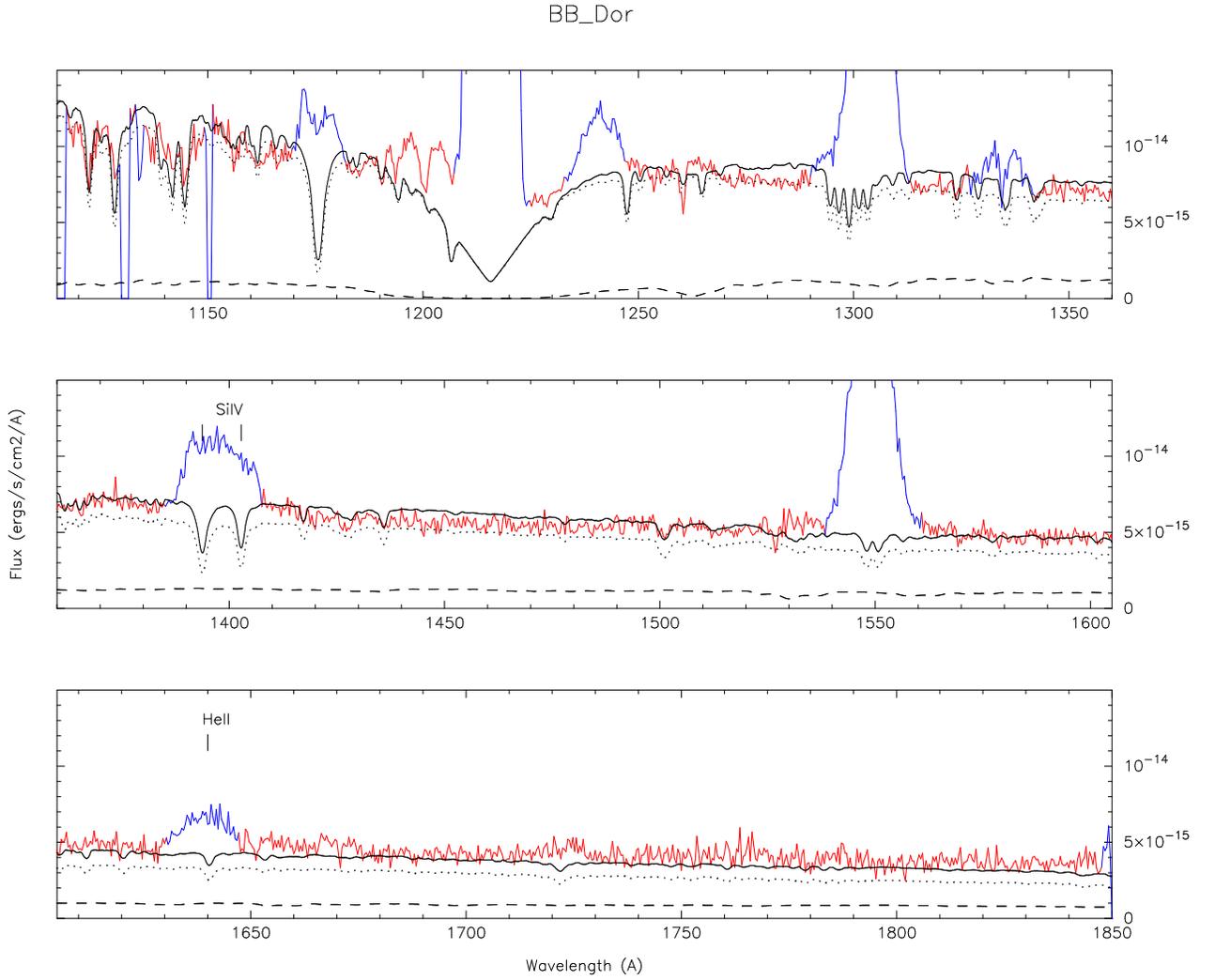}             
\figcaption{
Combined WD plus disk model to the COS spectrum of BB Dor. 
The WD has a temperature $T=35,000$ K, a projected
rotational velocity $V_{\rm rot} \sin{i} =200~$km$~$s$^{-1}$, 
and a mass of $0.8~M_{\odot}$. 
The disk has a mass accretion rate of 
$\dot{M}=3 \times 10^{-11}~M_{\odot}$yr$^{-1}$  
and an inclination $i=18^{\circ}$. 
The distance obtained is $d=$761 pc, and the reduced $\chi^2_{\nu}=1.23$.
The WD contributes 86\% of the FUV flux, and the disk contributes
the remaining 14\%. 
}
\end{figure}


\begin{thebibliography}{}

\bibitem[Andronov et al.(2003)]{and03}
Andronov, N., Pinsonneault, M., Sills, A.  2003, \apj, 582, 358 

\bibitem[Bisol et al.(2012)]{bis12}
Bisal, A.C., Godon, P., \& Sion, E.M. 2012, \pasp, 124, 158 

\bibitem[Chen et al.(2001)]{che01}
Chen, A., O'Donoghue, D., Stobie, R.S., Kilkenny, D., \& Warner, B. 2001,
\mnras, 325, 89 

\bibitem[Ellis et al.(1984)]{ell84} 
Ellis, G.L., Grayson, E.T., Bond, H.E. 1984, \pasp, 96, 283  

\bibitem[la Dous(1990)]{lad90} 
la Dous, C. 1990, in Cataclysmic Variables and Related Objects,
M. Hack (ed.), NASA/CNRS Monograph Series on Non-Thermal Phenomena
in Stellar Atmpsheres, p.14  

\bibitem[la Dous(1991)]{lad91} 
la Dous, C. 1991, \aap, 252, 100  

\bibitem[Godon et al.(2007)]{god07} 
Godon, P., Sion, E.M., Barrett, P.E., Szkody, P. 2007, 
\apj, 656, 1092 

\bibitem[Godon et al.(2008)]{god08} 
Godon, P., Sion, E.M., Barrett, P.E., Szkody, P., Schlegel, E.M. 2008, 
\apj, 687, 532 

\bibitem[Godon et al.(2006)]{god06} 
Godon, P., Sion, E.M., Cheng, F., Long, K.S., G\"ansicke, B.T., 
\& Szkody, P. 2006, \apj, 642, 1018  

\bibitem[Godon et al.(2012)]{god12}
Godon, P., Sion, E.M., Levay, K., Linnell, A.P., Szkody, P., Barrett, P.E.,
Hubeny, I., Blair, W.P. 2012, \apjs, 203, 29 

\bibitem[Hack \& la Dous(1993)]{hac93}
Hack, M., \& la Dous, C. 1993, 
{\it Cataclysmic Variables and Related Objects}, 
NASA SP-507, US Gov. Pringing Office, 
Washington DC, Chapter 3, p.95   


\bibitem[Hoard et al.(2004)]{hoa04}
Hoard, D.W., Linnell, A.P., Szkody, P., Fried, R.E., Sion, E.M., 
Hubeny, I., \& Wolfe, M.A. 2004,\apj, 604, 346  

\bibitem[Hodge et al.(2007)]{hod07}
Hodge, P.E., Keyes, C., Kaiser, M.E. 2007, AAS 211, 135.03, 
bulletin of the American Astronomical Society, vol.39, p.972 

\bibitem[Hodge(2011)]{hod11}
Hodge, P.E. 2011, in {\it Astronomical Data Analysis Software and Systems XX}   
eds. I.N. Evans, A. Accomazzi, D.J. Mink, \& A.H. Rots, (A.S.P.Conf.Series
442, ASP, San Fransisco), 391 

\bibitem[Howell et al.(2001)]{how01}
Howell, S., Nelson, L., \& Rappaport, S. 2001, \apj, 550, 897 

\bibitem[Hubeny(1988)]{hub88}
Hubeny, I. 1988, Comput. Phys. Commun., 52, 103

\bibitem[Hubeny \& Lanz(1995)]{hub95}
Hubeny, I.,\& Lanz, T. 1995, \apj, 439, 875

\bibitem[Knigge et al.(2011)]{kni11}
Knigge, C., Baraffe, I., \& Patterson, J. 2011, \apj, 194, 28 

\bibitem[Linnell et al.(2005)]{lin05}
Linnell, A.P., Szkody, P., G\"ansicke, B.T., Long, K.S., Sion, E.M., 
Hoard, D.W., \& Hubeny, I. 2005, \apj, 624, 923  

\bibitem[Linnell et al.(2007)]{lin07}
Linnell, A.P., Godon, P., Hubeny, I., Sion, E.M., Szkody, P. 2007, 
\apj, 662, 1204  

\bibitem[Linnell et al.(2008a)]{lin08a}
Linnell, A.P., Godon, P., Hubeny, I., Sion, E.M., Szkody, P.,
Barrett, P.E. 2008a, \apj, 676, 1447  

\bibitem[Linnell et al.(2008b)]{lin08b}
Linnell, A.P., Godon, P., Hubeny, I., Sion, E.M., Szkody, P. 2008b, 
\apj, 688, 568 

\bibitem[Linnell et al.(2009)]{lin09}
Linnell, A.P., Godon, P., Hubeny, I., Sion, E.M., Szkody, P.,
Barrett, P.E. 2009, \apj, 703, 1839  

\bibitem[Linnell et al.(2010)]{lin10}
Linnell, A.P., Godon, P., Hubeny, I., Sion, E.M., Szkody, P. 2010, 
\apj, 719, 271 

\bibitem[Mauche et al.(1988)]{mau88}
Maucher, C.W., Raymond, J.C., \& C\'ordova, F.A. 1988, \apj, 335, 829 

\bibitem[Nelson (2016)]{nel16}
Goliasch, J. \& Nelson, L. 2016, \apj, 809, 80 

\bibitem[Pala et al.(2016)]{pal16} 
Pala, A.F. et al. 2016, in prepration. 

\bibitem[Patterson(2002)]{pat02}
Patterson, J., 2002, Center for Backyard Astrophysics Communication
( http://cbastro.org/communications/news/messages/0278.html ) 

\bibitem[Press et al.(1992)]{numrec}
Press, W.H., Teukolsky, S.A., Vetterling, W.T., Flannery, B.P.,
Numerical Recipes in Fortran 77, The Art of Scientific Computing,
Second Edition, 1992, Cambridge University Press.

\bibitem[Pringle(1981)]{pri81} 
Pringle, J.E. 1981, \araa, 19, 137

\bibitem[Prinja et al.(2000)]{pri00}
Prinja, R.K., Ringwald, F.A., Wade, R.A., Knigge, C. 2000, \mnras, 312, 316  

\bibitem[Puebla et al.(2007)]{pue07}
Puebla, R.E., Diaz, M.P., Hubeny, I. 2007, \aj, 134, 1923 

\bibitem[Rappaport et al.(1982)]{rap82} 
Rappaport, S., Joss, P.C., \& Webbink, R.F. 1982, \apj, 254, 616  

\bibitem[Rappaport et al.(1983)]{rap83} 
Rappaport, S., Verbunt, F., \& Joss, P.C. 1983, \apj, 275, 713  

\bibitem[Starrfield (1971)]{sta71}
Starrfield, S. 1971, \mnras, 152, 307 

\bibitem[Rodr\'iguez-Gil et al.(2012)]{rod12} 
Rodr\'iguez-Gil, P., Schmidtobreick, L., Long, K.S., G\"ansicke, B.T., 
Torres, M.A.P., Rubio-D\'iez, M.M., \& Santander-Garc\'ia, M. 2012, 
\mnras, 422, 2332 

\bibitem[Rodr\'iguez-Gil et al.(2015)]{rod15} 
Rodr\'iguez-Gil, P., Shahbaz, T., Marsh, T.R. et al. 2015, \mnras, 452, 146 

\bibitem[Schmidtobreick et al.(2012)]{sch12} 
Schmidtobreick, L., Rodr\'iguez-Gil, P., Long, K.S., G\"ansicke, B.T., 
Tappert, C., \& Torres, M.A.P., 2012, \mnras, 422, 731  


\bibitem[Shakura \& Sunyaev(1973)]{sha73}
Shakura, N.I., \& Sunyaev, R.A. 1973, A\&A, 24, 337

\bibitem[Townsley \& Bildsten(2005)]{tow05}
Townsley, D.M., \& Bildsten, L. 2005, \apj, 628, 395 

\bibitem[Townsley \& G\"ansicke(2009)]{tow09}
Townsley, D.M., \& G\"ansicke, B.T. 2009, \apj, 693, 1007
     
\bibitem[Wade \& Hubeny(1998)]{wad98}
Wade, R.A., \& Hubeny, I. 1998, \apj, 509, 350

\bibitem[Warner(1995)]{war95}
Warner, B. 1995, Cataclysmic Variable Stars (Cambridge: Cambridge Univ.
Press)

\bibitem[Wood(1990)]{woo90}
Wood, M.A. 1990, Ph.D. thesis, University of Texas at Austin

\bibitem[Zellem et al.(2009)]{zel09}
Zellem, R., Hollon, N., Ballouz, R.-L., Sion, E.M., Godon, P., 
G\"ansicke, B.T., Knox, L. 2009, \pasp, 121, 942 

\bibitem[Zorotovic et al.(2011)]{zor11}
Zorotovic, M., Schreiber, M.R., G\"ansicke, B.T. 2011, \aap, 536, 42 

\end{thebibliography}
\end{document}